# Testing and Evaluation of Service Oriented Systems

Ashish Seth[#1], Himanshu Agarwal[#1], Ashim Raj Singla[*2]

[#1] *Punjabi University, Patiala, India*
[#2] *Indian Institute of Foreign Trade, New Delhi, India*
[1]ashish_may13@rediffmail.com
[1]himaggarwal@rediffmail.com
[2]arsngla@iift.ac.in

Abstract:
In Computer Science we have many established testing methods and tools to evaluate the software systems but unfortunately they don't work well for systems that are made up of services. For example, to users and systems integrators, services are just interfaces. This hinders white box testing methods based on code structure and data flow knowledge. Lack of access to source code also prevents classical mutation-testing approaches, which require seeding the code with errors. Therefore, evaluation of service oriented system has been a challenge, though there are large number of evaluation metrics exist but none of them is efficient to evaluate these systems effectively. This paper discusses the different testing tools and evaluation methods available for SOA and summarizes their limitation and support in context of service oriented architectures.

## 1. Introduction

Most organizations that want to build an SOA don't have a clue about how to approach the cost estimate. So, how to you calculate the cost of an SOA has been a challenge. We can't cost out an SOA like a construction project where every resource required is tangible and is easily accountable for calculating the total project costly. Since to compute the cost of many notions like : Understanding domain in proper context, understanding how much required resources cost, understanding how the work will get done and analyzing what can go wrong are some of intangible resources that are always required and are difficult to measure. According to D. Linthicum, the risk and impact of SOA are distributed and pervasive across applications, therefore, it is critical to perform an architecture evaluation early in the software life cycle [D. Linthicum, (2007)]. Because SOA involves the connectivity of multiple systems, business entities, and technologies, its overall complexity and the political forces involved need to be factored into architecture trade off considerations more than in single-application designs where technical concerns predominate.

SOA is a widely used architectural approach for constructing large distributed systems, which may integrate several systems that offer services and span multiple organizations. In this context, it is important that technical aspects be considered carefully at architectural design time. In a software architecture evaluation, we weigh the relevance of each design concern only after we understand the importance of each quality attribute requirement. Because decisions about SOA tend to be pervasive and have a significant and broad impact on business, therefore performing an early architecture evaluation is particularly valuable and is always recommended.

## 1.1. *Service Oriented Architecture (SOA)*

There are many definitions of SOA but none are universally accepted. What is central to all, however, is the notion of service. According to Phil B. *et al.(2007)*, in an SOA systems service is defined as follows.
- ☐ is selfcontained, highly modular and can be independently deployed.
- Is a distributed component and is available over the network and accessible through a name or locator other than the absolute network address.
- Has a published interface so the users of the service only need to see the interface and can be oblivious to implementation details.
- Stresses interoperability such that users and providers can use different implementation languages and platforms.
- Is discoverable, means users can look it up in a special directory service where all the services are registered.
- Is dynamically bound signifies that the service is located and bound at runtime. Therefore, service user does not need to have the service implementation available at build time.

These characteristics describe an ideal service. In reality, services implemented in service oriented systems lack or relax some of these characteristics, such as being discoverable and dynamically bound. Along with this there are some of the constraints that apply to the SOA architectural style are as follows [Phil B. *et al. (2007)*]
- Service users send requests to service providers.
- A service provider can also be a service user.
- A service user can dynamically discover service providers in a directory of services.
- An ESB can mediate the interaction between service users and service providers.

## 1.2. *Service*

Service is an implementation of a well-defined business functionality that operates independent of the state of any other service defined within the system. It has well-defined set of interfaces and operates through a pre-defined contract between the client of the service and the service itself, which must be dynamic, flexible for adding, removing or modifying services, according to business requirements. [Seth A, (2011)]. Services are loosely coupled, autonomous, reusable, and have well-defined, platform-independent interfaces, provides access to data, business processes and infrastructure, ideally in an asynchronous manner. Receive requests from any source making no assumptions as to the functional correctness of an incoming request. Services can be written today without knowing how it will be used in the future and may stand on its own or be part of a larger set of functions that constitute a larger service. Thus services within SOA
- Provides for a network discoverable and accessible interface
- Keeps units of work together that change together (high coupling)
- Builds separation between independent units (low coupling)

From a dynamic perspective, there are three fundamental concepts which are important to understand: the service must be visible to service providers and consumers, the clear interface for interaction between them is defined, and how the real world is affected from interaction between services. (See figure 1). These services should be loosely coupled and have minimum interdependency otherwise they can cause disruptions when any of services fails or changes.

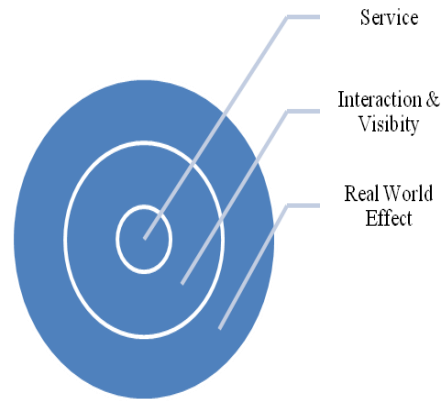

Figure 1. Service Model

### 1.3. *Enterprise Service Bus (ESB)*
An ESB is a flexible and standards based architecture that supports a wide array of transport mediums. Contrary to common belief, an ESB is not based solely on Web Services but based on the Enterprise Application Integration (EAI) pattern, thus, it is a standards-based integration platform that combines messaging, web services, data transformation and intelligent routing [Ahuja and Patel, (2011)]

Earlier model for integration like 'point to point' and 'spoke and wheel' had certain limitations. The complexity of application integration for a point to point model rises substantially with every new application that needs to communicate and share data with it. Every new application needs to have custom code written to 'glue' it to the existing network, and thus, increasing maintenance costs. This inefficient model gave rise to a new 'spoke and wheel' paradigm called the Enterprise Application Integration (EAI), in which, all communication is facilitated by the message broker. The message broker was designed not just for routing, but often used for data transformation as well. However, this architecture has scalability issues and introduces a single point of failure in the network.

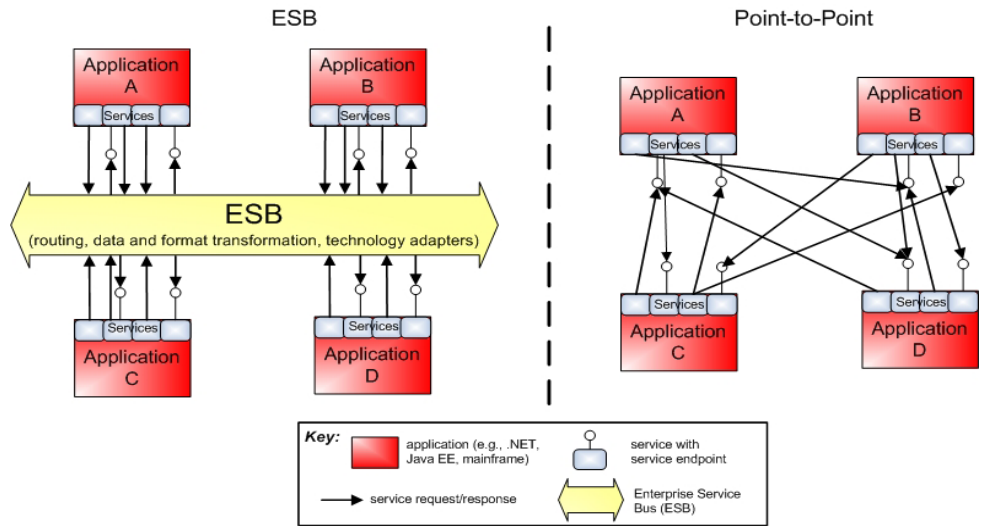

Figure 2: Comparison of ESB and Point-to-Point Integration Approaches [P. Bianco, 2007]

The Enterprise Service Bus is an improvement over these two architectures and plays a critical role in connecting heterogeneous applications and services in a Service-Oriented Architecture [Stojanovic, (2005)]. This middleware layer is responsible for not only transporting data, but also serves as a 'transformation' layer. This 'transformation' of data allows legacy systems to communicate and share data with newer applications.

## 2. Testing of Service Oriented Architectures

### 2.1. *ESB Evaluation factors*
Evaluating the cost and effectiveness of the SOA systems requires evaluation of ESB within the system. According to L. O'Brien, (2009) different factors were considered when comparing the open source ESBs. The following factors are suggested by different researches to determine performance and efficiency [L. O'Brien, (2009)].

*Mean Response Time*: One can calculated the Mean Response Time as the amount of time elapsed from the moment the request was sent to the time a reply was received.

*Throughput:* Throughput, as measured in transactions per second. A transaction was counted as successful, if it matched the expected response for the given request.

After retrieving the test data to compare the performances, we need a method to analyze the results. Simply calculating the throughput or the mean response times and generating graphs is not sufficient for the analysis.

### 2.2. *SOA Testing Dimensions and Roles*
Many established testing methods and tools to evaluate the software systems but unfortunately they don't work well for systems that are made up of services. For example, services are just interfaces to users and systems integrators. This hinders 'white box' testing methods based on code structure and data flow knowledge. Lack of access to source code also prevents classical mutation-testing approaches, which require seeding the code with errors. In this paper, we provide an overview of SOA testing are fundamental technical issues and comparative study of different solutions proposed, focusing on the SOA model designed for small and medium enterprises (SME's). Gerardo C and Massimiliano D discuss SOA testing across two dimensions [Gerardo and Massimiliano,(2006)]:

• *Testing perspectives*. Various stakeholders, such as service providers and end users, have different needs and raise different testing requirements.

• *Testing level*. Each SOA testing level, such as integration and regression testing, poses unique challenges.

Further, in order to understand the testing of service architecture completely, one needs to clear about the roles of services in different perspectives like service developer, service provider, service integrator, service user and third party certifier. Gerardo C. *et al.* (2006) describes the above terms as follows: (see table 1)

*Service developer*: the service developer tests the service to detect the maximum possible number of failures with an aim to release a highly reliable service.

*Service provider*: The service provider tests the service to ensure it can guarantee the requirements stipulated in the SLA with the consumer.

*Service integrator*: The service integrator test to gain confidence that any service to be bound to thier own composition fits the functional and nonfunctional assumptions made at design time.

*Third-party certifier*: The service integrator can use a third-party certifier to assess a service's fault-proneness.

*Service User*: only concern that the application he's using works while he's using it.

Regardless of the test method, testing a service-centric system requires the invocation of actual services on the provider's machine. This has several drawbacks. In most cases, service testing implies several service invocations, leading to unacceptably high costs and bandwidth use. [Gerardo and Massimiliano,(2006)].

3. **Related Work in cost evaluation for SOA systems**

3.1. *GQM method*
Since SOA follows different goals on different levels of EA abstraction, (Stephan A. *et. al.* 2009) shows that how these goals can be developed to metrics which can be consolidated in a measurement program. They present a method to design a set of metrics to measure the success of SOA. With these metrics the architects have a set of indicators showing the impact of each of their decisions during the process of building and maintaining SOA (see fig 2).

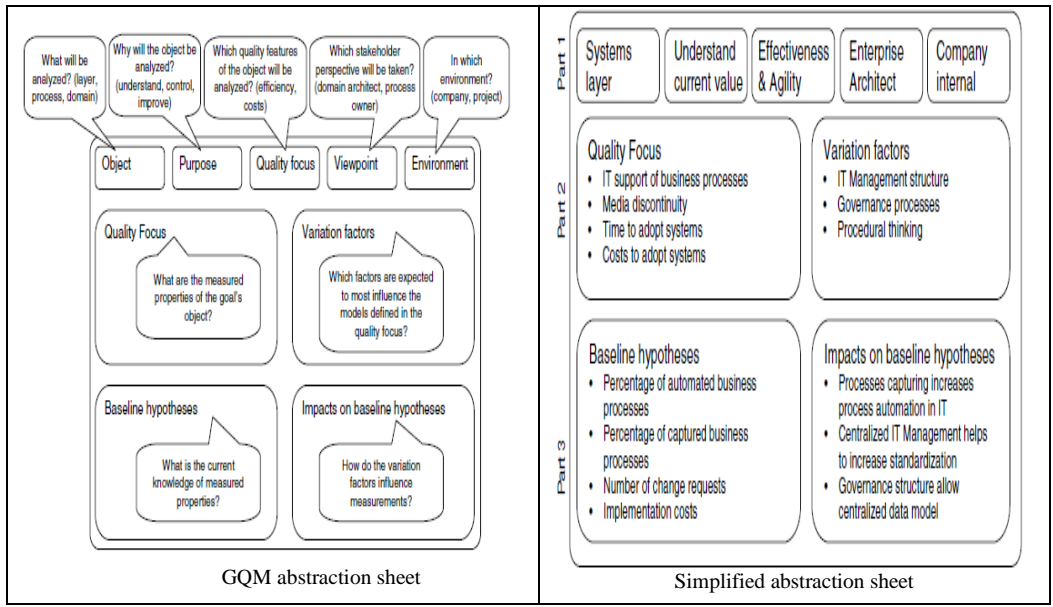

Figure 2. GQM Method [Van L.,*et al.,1998*]

Table 1. Highlights per testing dimension. Each stakeholder needs and responsibilities of are shown in black, advantages in green, issues and problems in red

| Testing levels | Testing perspectives | | | | |
| --- | --- | --- | --- | --- | --- |
| | Developer | Provider | Integrator | Third-party | User |
| Service functional testing | White-box testing available Service specification available to generate test cases Nonrepresentative inputs | Needs service specification to generate test cases Limited cost Black-box testing only Nonrepresentative inputs | Needs service specification Needs test suite deployed by service provider Black-box testing only High cost | Assesses only selected services and functionality on behalf of someone else (should be impartial assessment of all services) Small resource use for provider (one certifier tests the service instead of many ntegrators) Only nonrepresentative inputs High cost | Service-centric application self-testing to check that it ensures functionality during runtime Services have no interface to allow user testing |
| Service nonfunctional testing | Necessary to provide nonfunctional specifications to provider and consumers Limited cost Nonrealistic testing environment | Necessary to check own ability to meet SLA stipulated with user Limited cost Testing environment might not be realistic | High cost Might depend on network configuration Difficult to check whether SLA is met | Assesses performance on behalf of someone else Small resource use for provider (one certifier tests the service instead of many integrators) Nonrealistic testing environment High cost | Service-centric application self-testing to check that it ensures performance during runtime |
| Integration testing | Can be service integrator on his own | NA | Service call coupling increases because of dynamic binding Must regression test a composition after reconfiguration or rebinding Quality-of-service testing must consider all possible bindings | NA | NA |
| Regression testing | Limited cost (service can be tested off-line) Unaware of who uses the service | Limited cost (service can be tested off-line) Can be aware that the service has changed but unaware of how it changed | Might be unaware that the service has changed High cost | Retests service during its lifetime only on behalf of integrators, not other stakeholders High cost Lower-bandwidth use than having many integrators Nonrealistic regression test suite | Service-centric application self-testing to check that it works during evolution |

For most organizations, the first step of their SOA project is to figure out how much this SOA will cost. So that budget can be estimated to get the funding. The problem is that cost estimation of entire SOA components are not so easy and requires a clear understanding of the work that has to be done.

Dave Linthicum proposed a formula to figure out how much an SOA project will cost as follows [Dave Linthincum. 2011].

*Cost of SOA* = (Cost of Data Complexity + Cost of Service Complexity + Cost of Process Complexity + Enabling Technology Solution)

He further provide an example to arrive at the first variable, the cost of data complexity as follows:

*Cost of Data Complexity* = (((Number of Data Elements) x Complexity of the Data Storage Technology) x Labor Units)), where
- The "Number of Data Elements" is the number of semantics you're tracking in your domain, new or derived.
- Express the "Complexity of the Data Storage Technology" as a decimal between 0 and 1. (For instance, Relational is a .3, Object-Oriented is a .6, and ISAM is a .8.)
- "Labor Unit" is the amount of money it takes to understand and refine one data element. Dave said this could equal $100, for example.

As an example, you could arrive at a solution such as this:
*Cost of Data Complexity* = (((3,000) x .5) x $100) this equals $150,000 for this portion of your SOA costs.

Further, Dave suggested applying the same formulas to determine the costs of other variables, including Cost of Service Complexity, Cost of Process Complexity, and Enabling Technology Solution (which should be straightforward). Once you arrive at your Cost of SOA, Dave advises figuring in "10 to 20 percent variations in cost for the simple reason that we've not walked down this road before."

### 3.2. COCOMO II Related Approaches
COCOMO II (Constructive Cost Model) is one of the best-known and best-documented algorithmic models, which allows organizations to estimate cost, effort, and schedule when planning new software development activities. Tansey and Stroulia, (2010) have attempted to use COCOMO II to estimate the cost of creating and migrating services and suggested extension in COCOMO II to accommodate new characteristics of SOA based development. They also claimed that this model in general is inadequate to accommodate the cost estimation needs for SOA-based software development

Different survey and studies concluded that COCOMO II model by itself is inadequate to estimate effort required when reusing service-oriented resources. Although COCOMO II model has a large number of coefficients such as effort multipliers and scale factors, it is difficult to directly justify these coefficients in context of the cost estimation for SOA-based software development
.

### 3.3. *Functional Size Measurement Methods*

*3.3.1. IFPUG Function point Method*

It is obtained by summing up logical data groups and elementary processes classified respectively as Internal logical files, external interface files, external inputs, outputs or inquiries, with respect to the "application boundary", which separate the 'system' being measured from the user domain. IFPUG method provides a value adjustment factor (VAF) for taking into account several non-functional requirements for the final numerical assignments for the size of the systems being measured. Such factor does not include any specific consideration for software reuse resulting a function provided several to different systems is counted as many times, regardless of being designed and implemented only once or many times as well.

*3.3.2. COSMIC function point sizing method*

It's key concept are the possibility of viewing the system being measured as composed by different linked layers, by possibly separated software peer items within each layers, and the capability to specify different measurement viewpoints, based on different measurement purposes. Further more the COSMIC measure is more "associative" in mathematical sense than the IFPUG measure.

*3.3.3. Function Point Analysis and Software Sizing*

Size prediction for the constructed deliverables has been identified as one of the key elements in any software project estimation. SLOC (Source Line of Code) and Function Point are the two predominant sizing measures. Function Point measures software system size through quantifying the amount of functionality provided to the user in terms of the number of inputs, outputs, inquires, and files. Santillo, (2009) attempts to use the Function Point method to measure software size in an SOA environment. After comparing the effect of adopting the first and second generation methods (IFPUG and COSMIC respectively), Santillo identifies several critical issues. The prominent one is that SOA is functionally different from traditional software architectures, because the "function" of a service should represent a real-world self-contained business activity [G. Lewis *et al*. (2005)].

More issues appear when applying IFPUG to software system size measurement. Measuring with the COSMIC approach, on the contrary, is supposed to satisfy the typical sizing aspects of SOA-based software. However, there is a lack of guidelines for practical application of COSMIC measurement in SOA context. In addition to the application of Function Points, Liu *et al*. (2009) use Service Points to measure the size of SOA-based software. The software size estimation is based on the sum of the sizes of each service.

$$Size = (n,i) \Sigma ( Pi * P)$$

Where Pi is an infrastructure factor with empirical value that is related to the supporting infrastructure, technology and governance processes. P represents a single specific service's estimated size that varies with different service types, including existing service, service built from existing resources, and service built from scratch. This approach implies that the size of a service-oriented application depends significantly on the service type. However, the calculation of P for various services is not discussed in detail.

### 3.4. *SMAT-AUS Framework*
This framework reveals not only technical dimension but also social, cultural, and organizational dimensions of SOA implementation. When applying the SMAT-AUS framework to SOA-based software development, Service Mining, Service Development, Service Integration and SOA Application Development are classified as separate SOA project types. For each SOA project type, a set of methods, templates and cost models and functions are used to support the cost and effort estimation work for each project time which are then used to generate the overall cost of an SOA project (a combination of one or more of the project types).[ A. Bosworth, 2001]

*SMART Method* (Software Engineering Institute's Service Migration and Reuse Technique)
Except for the SMART (Software Engineering Institute's Service Migration and Reuse Technique) method [D. Linthicum, 2007] that can be adopted for service mining cost estimation, currently there are no other metrics suitable for the different projects beneath the SMAT-AUS framework. Instead, some abstract cost-estimation-discussions related to aforementioned project types can be found through a literature review. Umar and Zordan (2009) warn that both gradual and sudden migration would be expensive and risky so that costs and benefits must be carefully weighed. Bosworth (2010) gives a full consideration about complexity and cost when developing Web services. Liu et al. (2009) directly suggest that traditional methods can be used to estimate the cost of building services from scratch.

### 3.5. *Divide-and-Conquer Approach (D&C)*
The principle underlying D&C is to recursively decompose the problem into smaller sub-problems until all the sub-problems are sufficiently simple enough, and then to solve the sub-problems. Resulting solutions are then recomposed to form an overall solution. No mater where the D&C approach is applied the solution structure can be expressed explicitly in a program-like function such as:

$$\textit{Solution } x \equiv \quad \text{If IsBase (x)} \\ \text{Then SolveDirectly (x)} \\ \text{Else Compose (Solution(Decompose(x)))}$$

Where *x* is the original problem that will be solved through *Solution* procedure. *IsBase* is used to verify whether the problem *x* is primitive or not, which returns TRUE if *x* is a basic problem unit, or FALSE otherwise. *SolveDirectly* presents the conquer procedure. *Decompose* is referred to as the decomposing operation, while *Compose* is referred to as the composing operation [Zheng Li, Keung J, (2010)].

### 3.6. *Work Breakdown Structure (WBS) Approach*
This approach for cost estimation of SOA-based software is based on dealing separately with service parts. The WBS framework can help organizations simplify and regulate SOA implementation cost estimation by explicit identification of SOA-specific tasks in the WBS. Furthermore, both cost estimation modeling and software sizing work can be satisfied respectively by switching the corresponding metrics within this framework.

It is developed by starting with the end objective and successively re-dividing it into manageable components in terms of size, duration, and responsibility [T. Y. Lin, 2005]. In large projects, the approach is quite complex and can be as much as five or six levels deep.

Table2. Summary of different SOA based project evaluation approaches with the assumptions and limitations

| Approach | Solution Proposed | Assumptions | Limitation |
|---|---|---|---|
| Dave Linthicum formula | Cost of SOA = (Cost of Data Complexity + Cost of Service Complexity + Cost of Process Complexity + Enabling Technology Solution) | 10 to 20 percent variations in cost are expected. | • the other aspects of the calculation are suggested to follow similar means without clarifying essential matters<br>• this approach is not a real metric |
| COCOMO II Related Approaches | COCOMO II model has a large number of coefficients such as effort multipliers and scale factors<br>• . | COCOMO II considers two types of reused components, namely black-box components and white-box components. | • COCOMO II is generally inadequate to accommodate the cost estimation needs for SOA-based software development.<br>• COCOMO II model by itself is inadequate to estimate effort required when reusing service-oriented resources. |
| IFPUG | IFPUG provide Simple range matrices for software cost evaluation | IFPUG approach contributes to keep the method "simple and fast" | IFPUG measures leads to "same quantities" for "different" software units |
| COSMIC | COSMIC model provides open range scales to take into account possibly high complexity functions | COSMIC approach, is supposed to satisfy the typical sizing aspects of SOA-based software. | Wider set of guidelines for practical application of COSMIC measurement would still to test and experience. |
| Function Point Analysis and Software Sizing (based on IFPUG/COSMIC) | SLOC (Source Line of Code) and Function Point are the two predominant sizing measures | Function Point measures software system size through quantifying the amount of functionality provided to the user in terms of the number of inputs, outputs, inquires, and files | • When applying IFPUG to software system size measurement. For example, the effort of wrapping legacy code and data to work as services cannot be assigned to any functional size.<br>• there is a lack of guidelines for practical application of COSMIC measurement in SOA context. |
| Liu Service Points Method | Software size estimation is based on the sum of the sizes of each service.i.e Size = $(n,i) \Sigma ( Pi * P)$<br>where $Pi$ is an infrastructure factor with empirical value, is | • This approach implies that the size of a service-oriented application depends significantly on the service type.<br>• $P$ represents a single | The calculation of $P$ for various services is not discussed in detail. |

| | related to the supporting infrastructure, technology and governance processes. | specific service's estimated size that varies with different service types | |
|---|---|---|---|
| SMAT-AUS Framework | A generic SOA application could be sophisticated and comprise a combination of project types, breaking the problem into more manageable pieces (i.e. a combination of project types) | Entire SOA application is assumed to be classified as separate SOA project types into development, Service Mining, Service Development, Service Integration and SOA Application Development | Specifying how all of these pieces are estimated and the procedure required for practical estimation of software development cost for SOA-based systems is still being developed. |
| SMART (Software Engineering Institute's Service Migration and Reuse Technique) method [11] | can be adopted for service mining cost estimation | some abstract cost-estimation-discussions related to aforementioned project types can be found through a literature review. | Currently there are no other metrics suitable for the different projects beneath the SMAT-AUS framework. |
| GQM (goal/question/metrics) method | Based on the assumption that SOA follows different goals on different levels of EA abstraction | Assume that it is possible to identify certain SOA project types and certain context factors which can be combined to situations. | the identification of relevant project types and context factors are not clear. |
| Divide-and-Conquer (D&C) | It recursively decompose the problem into smaller sub problems until all the sub-problems are sufficiently simple enough, and then to solve the sub-problems. Resulting solutions are then recomposed to form an overall solution | • Assumed that the cost estimation for overall SOA-based software development can be separated into smaller areas with corresponding metrics.<br>• Approach mainly concentrating on cost estimation for Service Integration. | service classification can be different for different purposes, there is not a standard way to categorize services and method does not focus on this issue. |
| Work Breakdown Structure (WBS) approach | Based on the principle of Divide and Conquer theory, this framework can be helpful for simplifying the complexity of SOA cost estimation. | Through switching different type of metrics, this proposed framework could satisfy different requirements of SOA-based software cost estimation. | what will be the metric of different types is not properly explained |

## 4. Conclusion

Software cost estimation plays a vital role in software development projects, especially for SOA-based software development. However, current cost estimation approaches for SOA-based software are inadequate due to the architectural difference and the complexity of SOA applications. This paper discussed different testing and cost evaluation methods of service oriented systems. By using these techniques and identifying the support of each in context of service oriented systems can be helpful for simplifying the complexity of SOA cost estimation. By hosting different sets of metrics, this survey help not only for the complete cost estimation work but also for estimates the overall cost and effort through the independent estimation activities in different development areas of an SOA application.